\documentclass[twocolumn,aps,showpacs,prl]{revtex4}
\usepackage{graphicx}
\usepackage{epstopdf}

\begin{document}

\title{The layered compound CaClFeP is an Arsenic-free high $T_c$ iron-pnictide}

\author{Xun-Wang Yan$^{1,2}$}
\author{Zhong-Yi Lu$^{2}$}\email{zlu@ruc.edu.cn}

\date{\today}

\affiliation{$^{1}$Institute of Theoretical Physics, Chinese Academy
of Sciences, Beijing 100190, China }

\affiliation{$^{2}$Department of Physics, Renmin University of
China, Beijing 100872, China}

\begin{abstract}

We first analyze why the iron pnictides with high $T_c$
superconductivity so far are As-based, by the Hund's rule
correlation picture, then examine the P-based and Sb-based cases,
respectively. Consequently, we propose that CaClFeP with
ZrCuSiAs-type structure is an As-free high $T_c$ iron-pnictide. The
subsequent density functional theory calculations show that the
ground state of CaClFeP is of a collinearly antiferromagnetic order
on Fe moments with structural distortion, resulting from the
interplay between the strong nearest and next-nearest neighbor
antiferromagnetic superexchange interactions bridged by P atoms,
similar as the As-based pnictides. The other P-based pnictides are
either nonmagnetic or magnetic but with weak exchange interactions.
The Sb-based pnictides unlikely show high $T_c$ superconductivity
because of the existence of robust ferromagnetic order.

\end{abstract}

\pacs{74.70.Xa, 74.20.Pq, 74.20.Mn}

\maketitle


The discovery of high transition temperature $T_c$ superconductivity
in LaOFeAs by partial substitution of O with F atoms \cite{kamihara}
stimulates the intense studies on the iron pnictides. Upon doping or
high pressure, there are three types of iron pnictides reported to
show superconductivity, i.e. 1111-type $Re$OFeAs ($Re$ = rare earth)
\cite{kamihara} or BaFFeAs and SrFFeAs \cite{rotter1}, 122-type
$A$Fe$_2$As$_2$ ($A$=Ba, Sr, or Ca) \cite{rotter}, and 111-type
$B$FeAs ($B$ = alkali metal) \cite{wang}. All these pnictides share
the same structural feature that there exist the robust tetrahedral
layers where the Fe atoms are tetragonally coordinated by the As
atoms and the superconduction pairing may happen.

To our knowledge, all the iron pnictides showing high $T_c$
superconductivity so far are As-based. Here the high $T_c$ means
close to or above the theoretical maximum value of 40K predicted
from BCS theory\cite{WL} so that the corresponding superconductors
are considered unconventional. Up to now, the highest $T_c$ reported
in As-free pnictides is 17K shown in
(Fe$_2$P$_2$)(Sr$_4$Sc$_2$O$_6$) \cite{kamihara1}. Is it possible to
find a P- or Sb-based pnictide without As to show high $T_c$
superconductivity? Physically to answer this question essentially
depends upon how we understand the mechanism underlying the high
$T_c$ superconductivity in the pnictides. Likewise, solving this
question will greatly help us to understand the underlying
mechanism. Moreover, setting of this issue is very meaningful and
important in practice because Arsenic strong toxicity sets the
strict safety conditions to synthesize As-based pnictide samples,
which confine the samples making in a few research groups worldwide.
In this Letter, we show that CaClFeP with the ZrCuSiAs-type
structure is an As-free high $T_c$ pnictide with the similar
electronic structure and magnetic property as the As-based
pnictides.

It was universally found that all the As-based pnictides are in a
collinear antiferromagnetic (AFM) order below a
tetragonal-orthorhombic structural transition temperature
\cite{cruz,dong}. Accordingly, the superconducting pairing in the
As-based pnictides is now considered as being mediated by the spin
fluctuations. Regarding the mechanism behind the structural and AFM
transitions and their underlying relationship, there are basically
two contradictive views, based on the itinerant electron picture
\cite{mazin} and the local moment picture \cite{yildirim,si,ma1}
respectively. From the very beginning, we proposed \cite{ma1} the
fluctuating Fe local moments with the As-bridged AFM superexchange
interactions as the driving force upon the two transitions,
effectively described by the $J_1$-$J_2$ Heisenberg model. Here we
would like to emphasize the twofold meanings in our proposal
\cite{ma1}: (1) there are localized magnetic moments around Fe ions
and embedded in itinerant electrons in real space; (2) it is those
bands far from rather than nearby the Fermi energy that determine
the magnetic behavior of the As-based pnictides, namely the
hybridization of Fe with the neighbor As atoms plays a substantial
role. Here the formation of a local moment on Fe ion is mainly due
to the strong Hund's rule coupling on the Fe $3d$-orbitals, which is
about 0.6-0.8 eV/Fe\cite{ma1}. In this sense, our proposal can be
referred to as the Hund's rule correlation picture. We emphasize
again that the Arsenic atoms play a substantial role in our physical
picture. Based on this Hund's rule correlation picture, we had
successfully predicted \cite{ma} that the ground state of
$\alpha$-FeTe is in a bi-collinear AFM order, which was confirmed by
the later neutron scattering experiment \cite{shi}.

The main factor competing with the Hund's rule coupling is the
chemical bonding between the Fe $3d$ orbitals and As $4p$ orbitals.
Meanwhile the hybridization between the Fe $3d$ orbitals and As $4p$
orbitals will bridge the superexchange interactions between the Fe
magnetic moments induced by the Hund's rule coupling. It is thus a
dilemma to require strong superexchange interactions with robust Fe
magnetic moments. Chemically, the spin-degenerated Fe $3d$ orbitals
form stronger bonds with the As $4p$ orbitals than the
spin-polarized Fe $3d$ orbitals do. For the As-based pnictides,
inspection of the calculations \cite{ma1} shows that the spin
degeneration pushes down the energy levels raised by the Hund's rule
coupling, and then strengthens the bonding between the Fe $3d$
orbitals and As $4p$ orbitals, which makes substantial energy gain
to the nonmagnetic state. However, this is still not enough to
completely overcome the Hund's rule coupling, in need of an
additional energy gain of $\sim$0.2 eV/Fe, as suggested by
calculated total energy difference between the nonmagnetic state and
collinear AFM state \cite{ma1}. Consequently, there are magnetic
moments induced around Fe atoms with the AFM superexchange
interactions bridged by the As atoms.

In line of pnicogen atoms of P, As, and Sb, P has a much smaller
atomic radius with a larger electronegativity than As while Sb has a
much larger atomic radius with a smaller electronegativity than As.
Hence, it is very likely that in the most P-based pnictides there
will form a very strong bonding between P and Fe so that the Hund's
rule coupling is quenched. This is verified by the calculations
reported below that the most P-based pnictides are nonmagnetic. On
the other hand, in the most Sb-based pnictides there will form a
relatively weak bonding between Sb and Fe so that large moments are
induced around Fe atoms with robust ferromagnetic spin fluctuations,
which is also verified by the calculations reported below. In
contrast, Arsenic is in a proper balance to the dilemma. This
explains why the iron pnictides with high $T_c$ superconductivity,
discovered so far, are As-based rather than P-based or Sb-based. Now
we deduce that if the in-plane lattice parameters $a$ and $b$ of a
P-based (Sb-based) pnictide are more large (small), its electronic
structure and magnetic property are more similar to the ones of the
As-based pnictides since the P-Fe (Sb-Fe) bonding is more weakened
(strengthened). Referring to the As-based pnictides, we further
deduce that CaClFeP has the magnetic structure and property similar
to the ones of the As-based pnictides, in great possibility to
realize the high $T_c$ superconductivity.

\begin{figure}
\includegraphics[width=7.5cm]{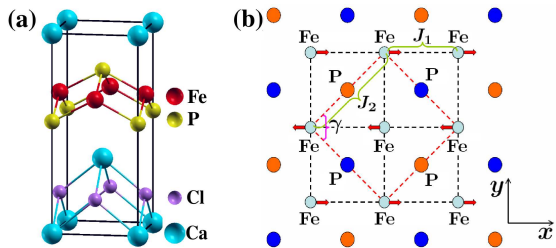}
\caption{(Color online) CaClFeP with the ZrCuSiAs-type structure:
(a) a tetragonal unit cell containing two formula units; (b)
schematic top view of the FeP layer. The small red dashed square is
an $a\times a$ unit cell, while the large black dashed square is a
$\sqrt{2}a\times \sqrt{2} a$ unit cell. The Fe spins in the
collinear antiferromagnetic order are shown by the red arrows.}
\label{figa}
\end{figure}

To verify the deduction, we carried out the density functional
theory (DFT) electronic structure calculations. In the calculations
the plane wave basis method was used \cite{pwscf}. We adopted the
generalized gradient approximation (GGA) of Perdew-Burke-Ernzerhof
\cite{pbe} for the exchange-correlation potentials. The ultrasoft
pseudopotentials \cite{vanderbilt} were used to model the
electron-ion interactions. After the full convergence test, the
kinetic energy cut-off and the charge density cut-off of the plane
wave basis were chosen to be 600eV and 4800eV, respectively. The
Gaussian broadening technique was used and a $k$-mesh of $16\times
16\times 8$ or $12\times12\times8$ was adopted to sample the
Brillouin zone of $a\times a\times c$ or $\sqrt{2}a\times
\sqrt{2}a\times c$ supercell, respectively. In the calculations, the
lattice parameters with the internal atomic coordinates were all
determined by the energy minimization. CaClFeP is in a tetragonal
layered structure with $P4/nmm$ symmetry. A crystal unit cell
consists of eight atoms with alternating FeP and CaCl layers along
the $c$ axis (Fig. \ref{figa}(a)).

\begin{figure}
\includegraphics[width=8.5cm]{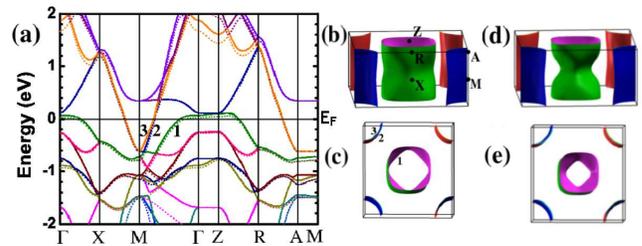}
\caption{(Color online) CaClFeP in the nonmagnetic state: (a) energy
band structure, in which solid lines for the undoping while dotted
lines for the 10\%-replacement of Ca by La; (b) the Fermi surface;
(c) top view on the Fermi surface. (d) and (e) correspond to the
case of 10\%-replacement of Ca by La.}\label{figb}
\end{figure}

As analyzed above, our calculations show that 111-type $B$FeP ($B$ =
Li or Na) and 122-type $A$Fe$_2$P$_2$ ($A$=Ba, Sr, or Ca) are all
nonmagnetic semimetals since the strong Fe-P bonding quenches the Fe
magnetic moments. Actually, the lattice parameters $a$ and $b$ of
these compounds are mainly determined by FeP layers, around 3.8 \AA
. The large parameters $a$ and $b$ go to the 1111-type, in which the
other layers strongly affect the lattice parameters. Indeed, we find
the large lattice parameters (a=b=4.14~\AA, c=8.70~\AA~) for CaClFeP
in magnetic phases.

In Fig. \ref{figb} we plot the electronic band structure and Fermi
surface of CaClFeP in the nonmagnetic state. There are three
cylinder-like Fermi surface sheets, among which the two are of
electron-type around M-A and the other one is of hole-type around
$\Gamma$-Z. The volumes enclosed by these Fermi sheets give 0.19
electrons/cell and equally 0.19 holes/cell, i.e. $1.31\times
10^{21}$/cm$^3$. The compound CaClFeP is thus a semimetal with a low
carrier density, between normal metals and semiconductors. The
density of states (DOS) at the Fermi energy is 3.65
states/(eV$\times$~unit cell). The corresponding electronic specific
heat coefficient $\gamma$ = $8.60~ mJ/(K^2\ast mol)$ and the Pauli
paramagnetic susceptibility $\chi_p$ = $1.49\times 10^{-9}~m^3/mol$.
The 10\%-replacement of Ca by La makes the electron-type and
hole-type Fermi sheets more matched in shape, as shown in Fig.
\ref{figb} (d) and (e).

In order to explore the magnetic structures and clarify the exchange
interactions between the Fe-Fe moments, we calculated three
different magnetic states, the ferromagnetic, checkerboard (Neel)
AFM, and collinear AFM states, which respective energies are
calculated to be (0.040, -0.044, -0.132) eV/Fe if the energy of the
nonmagnetic state is set to zero. The magnetic moment around each Fe
atom is found about $2.0 \mu_{B}$. All these magnetic states can be
self-consistently described by the following frustrated Heisenberg
model with the nearest and next-nearest neighbor couplings $J_1$ and
$J_2$,
\begin{equation}\label{eq:Heisenberg}
H=J_1\sum_{\langle ij \rangle}\vec{S}_i\cdot\vec{S}_j +J_2\sum_{ \ll
ij \gg}\vec{S}_i\cdot\vec{S}_j,
\end{equation}
whereas $\langle ij \rangle$ and $\ll ij \gg$ denote the summation
over the nearest and next-nearest neighbors, respectively. From the
calculated energy data, we find that $J_1 \sim 20.96~meV/S^2$ and
$J_2 \sim 32.58~meV/S^2$ (the detailed calculation is referred to
Appendix of Ref. \onlinecite{ma1}). The ground state of CaClFeP is
thus in a collinear AFM order.

\begin{figure}
\includegraphics[width=8.5cm]{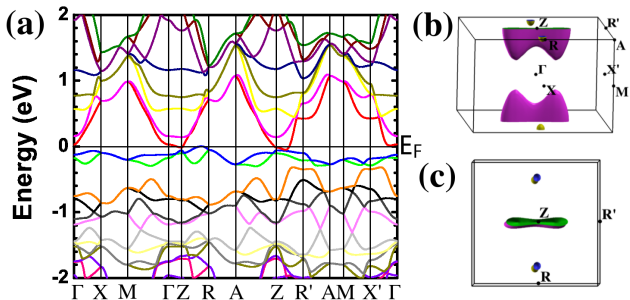}
\caption{(Color online) CaClFeP in the collinear-ordered
antiferromagnetic state with the in-plane angle
$\gamma=90.4^{\circ}$ (see Fig. 1(a)): (a) energy band structure;
(b) the Fermi surface; (c) top view on the Fermi surface. Notice
that $\Gamma X$ corresponds to the parallel-aligned moment line
while $\Gamma X^{\prime}$ corresponds to the antiparallel-aligned
moment line.} \label{figd}
\end{figure}

In the collinear AFM phase, a small structural distortion is found
to further make energy gain by 1.2 meV/Fe, which changes the angle
between two principal axes in ab-plane $\gamma$ from $90^\circ$ to
$90.4^{\circ}$ (see Fig. \ref{figa}(b)), equivalently, the lattice
unit cell slightly expands along the spin-antiparallel direction and
slightly shrinks along the spin-parallel direction. Here we
emphasize that our calculations show that the driving force upon
this tetragonal-orthorhombic structural distortion is nothing but
the magnetic interaction. More specifically it is the superexchange
interaction $J_2$ that induces the Fe-spin collinear AFM order and
drives the structural distortion to break the rotational symmetry,
as found in LaOFeAs \cite{ma1}. Moreover, the anti-parallel
alignment for the Fe moments between the neighboring FeP-layers was
found almost degenerate with the parallel alignment.

\begin{figure}
\includegraphics[width=6.0cm]{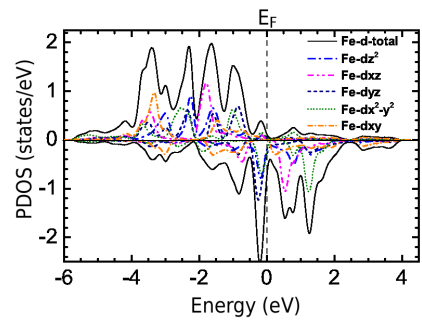}
\caption{(Color online) Calculated total and projected density of
states at the five Fe-$3d$ orbitals around one of the four Fe atoms
in the collinear antiferromagnetic state unit cell.}\label{figc}
\end{figure}

From Fig. \ref{figd}, we notice that the Fermi surface is more
distributed along $Z R^{\prime}$ or $\Gamma X^{\prime}$ than along
$Z R$ or $\Gamma X$ direction in Brillouin zone for CaClFeP in the
collinear AFM state, similar as in LaOFeAs and
BaFe$_2$As$_2$\cite{ma1,ma2}. This suggests that the
spin-antiparallel direction is more conductive than the
spin-parallel direction in real space. This anti-intuitive finding
was confirmed by the recent experiment on the detwinned
BaFe$_2$As$_2$ \cite{tanatar}.

Fig. \ref{figc} projects the density of states onto the five $3d$
orbitals of Fe in the collinear AFM of CaClFeP. We see that the five
up-spin orbitals are almost filled while the five down-spin orbitals
are nearly uniformly partially filled. This shows that the P
atom-imposed crystal field splitting is small. The formation of Fe
magnetic moments is thus due to the Hund's rule coupling, which is a
universal feature found in all the As-based pnictides
\cite{ma1,ma2}.

Other 1111-type P-based pnictides were also calculated. They all
show the similar electronic structures and magnetic properties but
with much smaller exchange couplings, as reported in Table
\ref{table}. These compounds have smaller lattice parameters (about
4 \AA ) than CaClFeP.

\begin{table}
\caption{Calculated exchange couplings $J_1$-$J_2$ (Fig.
\ref{figa}(b)) with measured superconductivity transition
temperature $T_c$.}\label{table}
\begin{tabular}{c c c c }
\hline
                  & $J_1$ (meV/Fe)& $J_2$ (meV/Fe)& $T_c$ (K) \\
\hline
LaOFeP                            & 0.52 &  7.83 & 7\cite{kamihara1} \\
(Fe$_2$P$_2$)(Sr$_4$Sc$_2$O$_6$)  & 3.74 & 12.24 & 17\cite{ogino} \\
BaFFeP                            & 0.62 & 11.20 & ? \\
CaClFeP                           &20.96 & 32.58 & ? \\
\hline
\end{tabular}
\end{table}

We have known that all the As-based pnictides have the similar
electronic structures and magnetic properties. Upon doping or high
pressure, the AFM long-range order is suppressed. The remanent AFM
fluctuations are now considered to be responsible for the
superconducting pairing in pnictides \cite{ma1}, similar as in
cuprate superconductors. We had systematically calculated the
exchange couplings $J_1$ and $J_2$ for the different As-based
pnictides. And we found that phenomenologically the
superconductivity transition temperature $T_c$ is proportional to
$J_2$\cite{ma2}. Theoretically, the two-band $t$-$J_1$-$J_2$ model
also indicates that the $T_c$ in pnictides is proportional to the
$J_2$\cite{seo}.

As we have just shown, CaClFeP takes the similar electronic
structure and magnetic property as the As-based pnictides and the
exchange couplings $J_1$ and $J_2$ of it are as large as the ones of
BaFe$_2$As$_2$ \cite{ma2}. We thus predict that the high $T_c$
(likely over 40K) superconductivity can be realized in CaClFeP.
Meanwhile, for LaOFeP and (Fe$_2$P$_2$)(Sr$_4$Sc$_2$O$_6$) the
smallness of calculated $J_1$ and $J_2$ is consistent with the
observed low $T_c$ (Table \ref{table}).

\begin{figure}
\includegraphics[width=8.5cm]{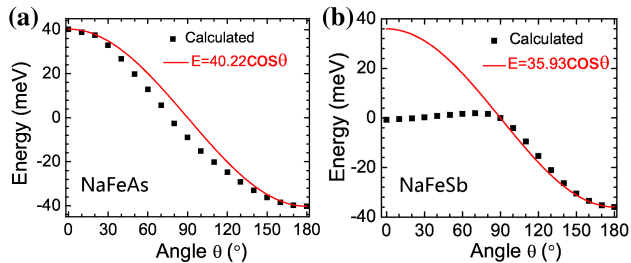}
\caption{(Color online) Calculated magnetic energy versus the angle
$\theta$ between the two sublattice spin orientations, from the
ferromagnetic order ($\theta=0^{\circ}$) to the Neel AFM order
($\theta=180^{\circ}$). (a) NaFeAs and (b) NaFeSb. }\label{fige}
\end{figure}

We have performed the DFT calculations on Sb-based pnictides,
including 111-type, 122-type, and 1111-type compounds. We find that
all these Sb-based pnictides are similar with each other in
electronic structures and magnetic properties. In magnetic states,
they show rich magnetic orders like the checkerboard AFM,
bi-collinear AFM, and collinear AFM orders, however, there are much
larger moments formed at Fe atoms ($\sim 3\mu_B$/Fe) in the Sb-based
pnictides than in the As-based pnictides because of the strong
Hund's rule effect and the weak Fe-Sb bonding, as analyzed above.
Besides, even though the collinear AFM order is the most stable
energetically, there exists a stable ferromagnetic order in the
Sb-based pnictides, independent of the AFM fluctuations. This is
essentially different from the As-based pnictides, in which a
ferromagnetic order is never stable.

Fig. \ref{fige}(a) shows that for NaFeAs the ferromagnetic order is
a maximum energetically, and the magnetic energy versus the angle
$\theta$ between the two sublattice spin orientations is well
described by the nearest neighbor exchange Heisenberg model. The
calculations also show that CaClFeP behaves likewise. In contrast,
for NaFeSb the ferromagnetic state is a local minimum in the energy
landscape, lower than the nonmagnetic state (Fig. \ref{fige}(b)).
And the variation of the magnetic energy with the angle $\theta$
severely deviates from the exchange Heisenberg model. This shows
that the ferromagnetic state has a different exchange interaction
origin than the AFM states, which is likely related to the possible
orbital ordering. This ferromagnetic state can be stable until
applying a high pressure of 25 GPa. Since such ferromagnetic
correlations are against the superconducting pairing induced by the
AFM fluctuations, we don't think there is the high $T_c$
superconductivity realized in Sb-based pnictides. To our knowledge,
no Sb-based pnictides are reported to show high $T_c$
superconductivity in experiment.

In summary, by the Hund's rule correlation picture we may well
understand the iron pnictides, in which the different magnetic
structures and properties result mainly from the competition between
the Hund's rule effect on Fe 3d-orbitals and the Fe-P, Fe-As, or
Fe-Sb bonding. In search for an As-free high $T_c$ pnictide, we show
that CaClFeP is an AFM semimetal with the strong P-bridged nearest
and next-nearest neighbor AFM superexchange interactions, which give
rise to the collinear AFM order on Fe moments in the ground state,
similar as LaOFeAs and BaFe$_2$As$_2$. Upon doping or high pressure,
CaClFeP is predicted to show high $T_c$ superconductivity. The other
1111-type P-based iron-pnictides are also AFM semimetals but with
weak exchange interactions, while both 111-type and 122-type P-based
iron-pnicitdes are nonmagnetic semimetals. We also show that the
Sb-based pnictides unlikely show high $T_c$ superconductivity.


This work is supported by National Natural Science Foundation of
China and by National Program for Basic Research of MOST, China.


\begin{references}

\bibitem{kamihara}Y. Kamihara, {\it et al.}, J. Am. Chem. Soc. {\bf 130}, 3296
(2008).
\bibitem{rotter1}M. Tegel, {\it et al.}, Europhys. Lett. {\bf 84}, 67007
(2008); X. Zhu, {\it et al.}, Europhys. Lett. {\bf 85}, 17011
(2009).
\bibitem{rotter}M. Rotter {\it et al.}, Phys. Rev. Lett. {\bf 101}, 107006 (2008).
\bibitem{wang}X. C. Wang {\it et al.}, Solid State Commun. {\bf 148}, 538 (2008).
\bibitem{WL} W.L. McMillan, Phys. Rev. {\bf 167}, 331 (1968).
\bibitem{kamihara1}Y. Kamihara, {\it et al.},
J. Am. Chem. Soc. {\bf 128}, 10012 (2006).
\bibitem{cruz} C. de la Cruz {\it et al.}, Nature (London) {\bf 453}, 899 (2008).
\bibitem{dong} J. Dong {\it et al.}, Europhys. Lett. {\bf 83}, 27006 (2008).
\bibitem{mazin}I.I. Mazin, {\it et al.},
Phys. Rev. Lett. {\bf 101}, 057003 (2008).
\bibitem{yildirim} T. Yildirim, Phys. Rev. Lett. {\bf 101}, 057010
(2008).
\bibitem{si} Q. Si and E. Abrahams, Phys. Rev. Lett. {\bf 101}, 076401
(2008).
\bibitem{ma1}F. Ma, Z.Y. Lu, and T. Xiang, Phys. Rev. B {\bf 78}, 224517
(2008).
\bibitem{ma} F. Ma {\it et al.}, Phys. Rev. Lett. {\bf 102}, 177003 (2009).
\bibitem{shi} S.L. Li, {\it et al.}, Phys. Rev. B {\bf 79}, 054503 (2009).
\bibitem{pwscf}P. Giannozzi {\it et al.}, http://www.quantum-espresso.org.
\bibitem{pbe} J.P. Perdew, K. Burke and M. Ernzerhof, Phys. Rev. Lett. {\bf 77} 3865
(1996).
\bibitem{vanderbilt}D. Vanderbilt, Phys. Rev. B {\bf 41}, 7892 (1990).
\bibitem{ma2} F. Ma, Z.Y. Lu, and T. Xiang, Front. Phys. China, {\bf 5(2)}, 150
(2010).
\bibitem{tanatar} M.A. Tanatar, {\it et al.}, Phys. Rev. B {\bf 81}, 184508 (2010).
\bibitem{seo} K. Seo, A. B. Bernevig, J. Hu, Phys. Rev. Lett. {\bf 101}, 206404 (2008).
\bibitem{ogino}H. Ogino, {\it et al.},
Supercond. Sci. Technol. {\bf 22}, 075008 (2009).

\end{references}
\end{document}